\documentclass[twocolumn,prl,showpacs]{revtex4}

\usepackage{graphicx}
\usepackage{dcolumn}
\usepackage{float}
\usepackage{amsmath}

\newcommand{\comment}[1]{}

\begin{document}

%\preprint{version 0}

\title{Phonon spectroscopy through the electronic density of states in graphene}

\author{E.J. Nicol$^*$ and J.P. Carbotte$^\dagger$}
\affiliation{$^*$Department of Physics, University of Guelph,
Guelph, Ontario, Canada, N1G 2W1}
\affiliation{$^\dagger$Department of Physics and Astronomy, McMaster University,
Hamilton, Ontario, Canada, L8S 4M1} 
\affiliation{$^\dagger$The Canadian
Institute for Advanced Research, Toronto, Ontario, Canada, M5G 1Z8}

\date{\today}

\begin{abstract}
{We study how phonon structure manifests itself in the electronic
density of states of graphene. A procedure for extracting the value of
the electron-phonon renormalization $\lambda$ is developed. In
addition, we identify direct phonon structures. With increasing
doping, these structures, along with $\lambda$, grow in amplitude and
no
longer display particle-hole symmetry.
}
\end{abstract}


\maketitle
% body of paper here

It is a remarkable result of many body physics 
that in electronic systems in which
the density of states (DOS) is constant on the scale of a phonon
energy, electron-phonon renormalizations entirely drop out\cite{Grimvall,Prange}
 and no
phonon signatures are expected or seen. This has now changed with
the advent of graphene. Graphene was isolated only in 2004\cite{geimscience,novoselov} but
has since been extensively studied and found to exhibit many 
unusual properties. These include a novel quantum Hall effect, a
minimum conductivity, a Berry phase of $\pi$ and other effects related
to the chirality of its charge carriers.\cite{geimnature,kimnature,geimnatmat} 

In this letter, we show how
measurements of the DOS in graphene offer a new opportunity
to obtain detailed information on electron-phonon coupling
in sharp contrast to the case of ordinary metals. This is important
as it  illustrates
a strong violation of a well-established result of many body physics
and provides unusual doping-dependent predictions for the manifestation
of the electron-phonon interaction in experiments on graphene. This
result arises
for two reasons. 
First, the charge carriers exhibit relativistic dispersions
with quasiparticle energy $(\epsilon_{\bf k})$ linear, rather
than quadratic, in momentum (${\bf k}$), $\epsilon_{\bf k}=\pm\hbar
v_0|{\bf k}|$. Here, $v_0$ plays the role of the velocity of light
and the $\pm$   gives the upper and lower Dirac cones, respectively.
At neutrality, the lower cone is fully occupied and the upper one,
empty. 
This dispersion gives rise to an energy dependence of the DOS
which is linear to high energies.
Second, both theory and experiment indicate
that the major coupling is to high energy phonons of order
200 meV.\cite{Park,li,bostwick} Thus, an electron scattering
from an initial state to a final state through the assistance of a phonon
will sample changes in the DOS on the scale of the phonon energy,
which is significant in graphene.
Another special feature of graphene is that 
the number of charge carriers can be changed
through charging in a field effect device where the chemical potential
($\mu_0$), measured with respect to the Dirac point (neutrality
point), is proportional to the square root of the gate voltage.
These characteristics offer a rich new spectroscopy for the study
of phonon effects including their variation with $\mu_0$.

In graphene, the bare band
DOS 
varies linearly with energy and hence the renormalized density of states
$N(\omega)$
is given by\cite{knigavko,frank,boza,anton} 
\begin{equation}
\frac{N(\omega)}{N_\circ}=\int_{-W_C}^{W_C}\!\!\! d\epsilon \frac{|\epsilon|}{\pi}
\frac{-{\rm Im}\Sigma(\omega)}
{[\omega-{\rm Re}\Sigma(\omega)+\mu-\epsilon]^2+[{\rm Im}\Sigma(\omega)]^2},
\label{eq:dos}
\end{equation}
with $N_\circ = 2/\pi\hbar^2v_0^2$. 
In Eq.~(\ref{eq:dos}), $W_C$ is an upper cutoff on the Dirac cones 
given by $\sqrt{\pi\sqrt{3}} t$, 
with $t$ the nearest neighbor hopping parameter, 
and $\Sigma(\omega)$ is the electronic
self-energy given by\cite{frank} 
\begin{eqnarray}
\Sigma(\omega)&=&\int_0^{+\infty}d\nu
\alpha^2F(\nu)\int_{-\infty}^{+\infty}
 d\omega'\frac{N(\omega')}{N_\circ W_C}\nonumber\\
&&\times\biggl[\frac{n(\nu)+f(-\omega')}{\omega-\omega'-\nu+i0^+}
+\frac{n(\nu)+f(\omega')}{\omega-\omega'+\nu+i0^+}\biggr] ,
\label{eq:Sigma}
\end{eqnarray}
where $\alpha^2F(\nu)$ is the electron-phonon spectral density
and $n(\nu)$ and $f(\omega')$ are, respectively, the Bose-Einstein and 
Fermi-Dirac distribution functions at temperature $T$.
For the bare band case $(\Sigma(\omega)\to 0)$, 
the chemical potential $\mu$ reduces to its
noninteracting value $\mu_0$ and sets the doping level. Also, 
the Lorentzian form
in Eq.~(\ref{eq:dos}) reduces to $\delta(\omega+\mu_0-\epsilon)$
and the DOS becomes $|\omega+\mu_0|$. 

Park et al.\cite{Park}
performed a full first-principles study of the electron-phonon interaction
in graphene and found that the result could be approximated by an
Einstein mode at 200 meV. For coupling to an Einstein mode, the electronic
self-energy in a system with a linear DOS
can be evaluated in the usual manner to give, at zero temperature,
an analytic form with\cite{stauber}
\begin{eqnarray}
{\rm Re}\Sigma(\omega)&=&\displaystyle
\frac{A}{W_C}\Biggl\{\omega_E \ln\Biggl|
\displaystyle
\frac{(\mu_0+\omega+\omega_E)^2}{(\omega^2-\omega_E^2)}\Biggr|\nonumber\\
&-&(\mu_0+\omega)\ln\Biggl|
\displaystyle
\frac{W_C^2(\omega+\omega_E)}
{(\omega-\omega_E)(\omega+\mu_0+\omega_E)^2}\biggr|\Biggr\},
\label{eq:resigma}
\end{eqnarray}
where $A$ is the area under the Einstein mode
and $\omega_E$ is the Einstein frequency. For simplicity, we have assumed
in writing Eq.~(\ref{eq:resigma}) that $W_C$
is larger than any other energy of interest, but in all numerical results
presented here, this approximation was not made.
The corresponding imaginary part is 
$
-{\rm Im}\Sigma(\omega)=\frac{\pi A}{W_C}|\omega-\omega_E+\mu_0|$, for
$\omega_E<\omega<W_C-\mu_0+\omega_E$ and 
$\frac{\pi A}{W_C}|\omega+\omega_E+\mu_0|$ for$ -\omega_E>\omega>-W_C-\mu_0-\omega_E$.
In terms of this self-energy, 
the renormalized density of states is given by 
\begin{widetext}
\begin{equation}
\frac{N(\omega)}{N_\circ}=\frac{\tilde\omega}{\pi}\biggl[
2\tan^{-1}\biggl(\frac{\tilde\omega}{\Gamma}\biggr)
-\tan^{-1}\biggl(\frac{\tilde\omega-W_C}{\Gamma}\biggr)
-\tan^{-1}\biggl(\frac{\tilde\omega+W_C}{\Gamma}\biggr)
\biggr]
\displaystyle
+\frac{\Gamma}{2\pi}\ln\biggl(\frac{[(\tilde\omega-W_C)^2+\Gamma^2]
[(\tilde\omega+W_C)^2+\Gamma^2]}{(\tilde\omega^2+\Gamma^2)^2}\biggr),
\label{eq:dosformula}
\end{equation}
\end{widetext}
where $\Gamma=-{\rm Im}\Sigma(\omega)$ and $\tilde\omega=\omega-{\rm
  Re}\Sigma(\omega)+\mu$.
For
finite $\mu$, the problem no longer has particle-hole symmetry and 
${\rm Re}\Sigma(\omega=0)$ is not zero and provides a
shift in chemical potential from bare to dressed value with
$\mu=\mu_0+{\rm Re}\Sigma(\omega=0)$.\cite{boza,luttinger}
 For a clean system $\Gamma$ will vanish for 
$-\omega_E<\omega<\omega_E$, and  Eq.~(\ref{eq:dosformula})
reduces
to
\begin{equation}
\frac{N(\omega)}{N_0}=\tilde\omega {\rm sgn}\tilde\omega, \quad {\rm
  for} -\omega_E<\omega<\omega_E.
\label{eq:renormdos}
\end{equation}
In this special range, the DOS is very closely related to the ${\rm Re}\Sigma(\omega)$. Returning to Eq.~(\ref{eq:dos}), it is important
to realize that, for infinite bands with constant DOS, the $|\epsilon|$
factor would  not appear and the integral over $\epsilon$ would give a constant
independent of $\omega$ so that phonon renormalizations simply drop out.
Graphene is very different.

\begin{figure}[ht]
\begin{picture}(250,200)
\leavevmode\centering\includegraphics{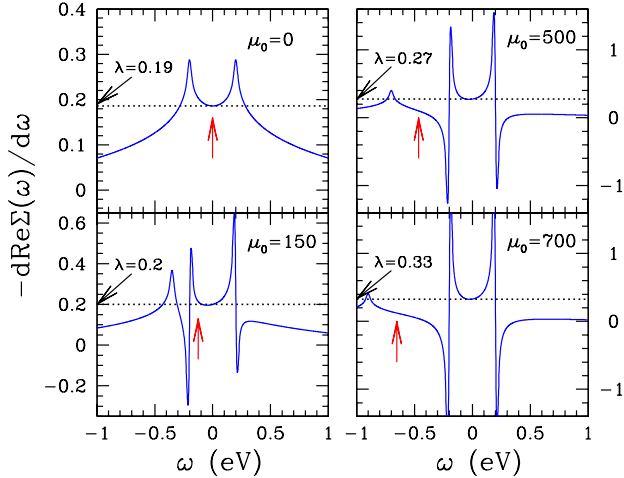}
\end{picture}
\vskip -30pt
\caption{(Color online)  $-d{\rm Re}\Sigma(\omega)/d\omega$
  vs $\omega$ for a truncated Lorentzian electron-phonon spectral
  density
peaked around $\omega_E=200$ meV. Shown
are curves for
 $\mu_0=$ 0, 150, 500, 
and 700 meV.
}
\label{fig1}
\end{figure}

While Eq.~(\ref{eq:resigma}) has been written for a single
Einstein oscillator, it nevertheless
provides us with valuable insight into the 
relationship between phonon structure and the real part of the
self-energy.
 ${\rm Re}\Sigma(\omega)$ has singularities of the
form
$\ln|\omega\pm\omega_E|$ at $\omega=\pm\omega_E$ and a third weaker
logarithmic
singularity of the type
$(\omega+\omega_E+\mu_0)\ln|\mu_0+\omega+\omega_E|$
at $\omega=-(\mu_0+\omega_E)$. The neutrality point is special,
however.
For $\mu_0=0$, only two singularities remain and they are both of the
weaker
kind $(\omega\pm\omega_E)\ln|\omega\pm\omega_E|$. In a real system
there will of course be a distribution of phonons and the self-energy of
Eq.~(\ref{eq:resigma}) needs to be averaged over such a
distribution. This will reduce the prominence of the expected
singularities in
this quantity. In such a case it becomes useful
to consider a first derivative $-d{\rm
  Re}\Sigma(\omega)/d\omega$. This is shown in Fig.~\ref{fig1} for
four values of the chemical potential, $\mu_0=0, 150, 500$ and 700 meV.
The
phonon
distribution used in these numerical calculations was a truncated
Lorentzian
centered around $\omega_E=200$ meV with width $\delta=15$ meV.\cite{frank,longpaper}. 
As expected the top left frame exhibits
only
two phonon anomalies while the three other frames have three. Also in
these
three cases the anomalies at $\omega=\pm\omega_E$ are much more pronounced
than the ones at $\omega=-(\omega_E+\mu_0)$, and also than
those in the top left frame. In all four frames, the black dotted
horizontal line was drawn through the local minimum at $\omega=0$ and
identifies the value of the electron-phonon mass renormalization
parameter
$\lambda$ as we will now describe. For $\omega$ small near the Fermi
energy $(\omega=0)$, 
${\rm Re}\Sigma(\omega)$ in Eq.~(\ref{eq:resigma}) can be shown
to
vary as  ${\rm Re}\Sigma(\omega)\simeq -\lambda\omega+{\rm
  Re}\Sigma(\omega=0)$
and the dressed quasiparticle energy $E_{\bf k}$ is given by the
equation
$E_{\bf k}-{\rm Re}\Sigma(E_{\bf k})+\mu=\pm\hbar v_0|{\bf k}|=E_{\bf
  k}(1+\lambda)+\mu_0$. Or $E_{\bf k}=[\pm\hbar v_0|{\bf
    k}|-\mu_0]/(1+\lambda)=\pm\hbar v_0(k-k_F)/(1+\lambda)$, which
means that $\lambda$ simply renormalizes the bare Fermi velocity from 
$v_0$ to $v_0^*\equiv v_0/(1+\lambda)$. As Fig.~\ref{fig1}
shows, $\lambda$ grows with increasing $\mu_0$ as
Eq.~(\ref{eq:resigma})
implies. The red arrow indicates the Dirac point defined by $|{\bf k}|=0$.

\begin{figure}[ht]
\begin{picture}(250,200)
\leavevmode\centering
\includegraphics{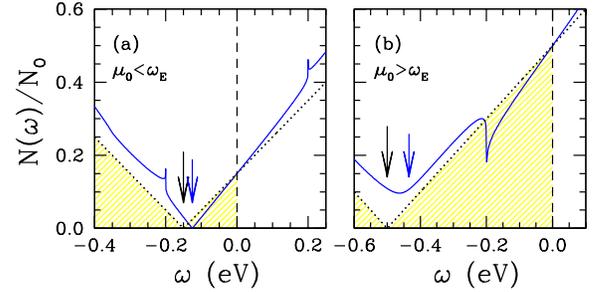}
\end{picture}
\vskip -100pt
\caption{(Color online) (a) $N(\omega)/N_0$ (in eV) vs $\omega$
for the bare chemical potential $\mu_0=150$ meV. The solid blue curve
gives the phonon renormalized case and the black dotted, the bare
band case.
The arrows show the bare (long black) and renormalized (short blue)
value of $\mu$. (b) Same as for (a) but with $\mu_0=500$ meV.
}
\label{fig2}
\end{figure}

The relationship  between boson structure in the self-energy and its
manifestation
in the DOS is given by Eq.~(\ref{eq:dosformula}). Results for
$N(\omega)$
are shown in Fig.~\ref{fig2}. The frame (a) is for $\mu_0=150$ meV,
which is smaller than $\omega_E$ and (b) is for $\mu_0=500$ meV
$>\omega_E$.
The shaded yellow region is the occupied part of the bare band 
which is shown as the black
dotted
curve. Phonon renormalizations change the shape of the DOS and hence
the value of the chemical potential must be altered to keep the
correct
number of particles. The long black and short blue arrows point to the
value
of the bare and dressed chemical potential with $\mu-\mu_0={\rm
  Re}\Sigma(\omega=0)$. Phonon anomalies
are clearly seen in the dressed curves. To  emphasize this structure
an Einstein spectrum was used with $\omega_E=200$ meV so that the
phonon
structures fall at $\omega=\pm\omega_E$, one on either side of the
Fermi 
energy. The expected singularity at $\omega=-(\omega_E+\mu_0)$
is by comparison very weak and appears as a slight change in
slope in Fig.~\ref{fig2}. Two
additional features of these curves are to be noted. At the Fermi
energy ($\omega=0)$, 
the dressed and bare DOS have exactly the same value. In the
region
of the Fermi energy Eq.~(\ref{eq:renormdos})
applies and $N(\omega)/N_0=|\omega(1+\lambda)+\mu_0|$,
which differs from its bare value only by the additional factor of 
$(1+\lambda)$. At $\omega=0$, this difference disappears and
dressed
and bare DOS are the same. Phonons do not change the value of the
DOS at the Fermi level.
The slope out of $\omega=0$, however, is
changed by a factor of $(1+\lambda)$ as can be seen in both frames
of Fig.~\ref{fig2} and we also note that this linear behavior persists
over a considerable energy range set by the value of the Einstein
oscillator.
Recognizing that the normalization for the DOS is $N_0\sim 1/v_0^2$, one might
naively think that the $(1+\lambda)$ renormalization can be included
in $N(\omega)$ simply by changing $v_0$ to $v_0^*$ in $N_0$, but we see
here
that this is not correct. Only one $(1+\lambda)$ factor enters and not
its
square. The basic reason underlying this fact is that the coherent
part of the electronic Green's function, which defines the quasiparticles
in the interacting system, contains only $1/(1+\lambda)$ of the spectral weight.
The remainder $\lambda/(1+\lambda)$ is found in the incoherent
piece describing phonon-assisted processes.

\begin{figure}[ht]
\begin{picture}(250,200)
\leavevmode\centering\includegraphics{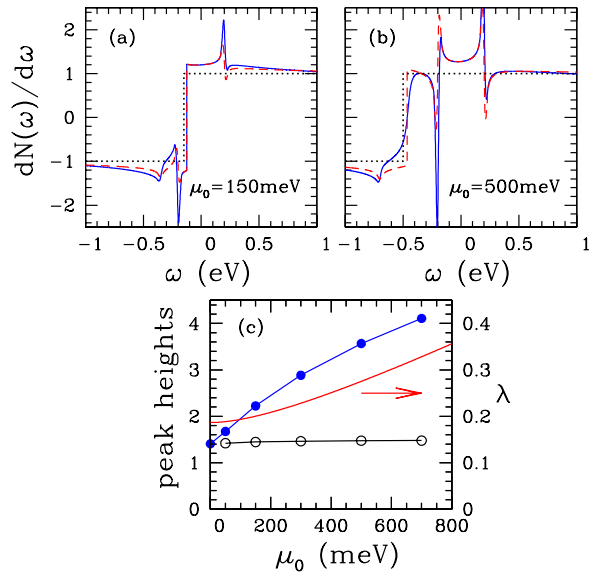}
\end{picture}
\vskip 10pt
\caption{(Color online) (a) $dN(\omega)/d\omega$
vs $\omega$ (solid blue curve) for $\mu_0=150$ meV. 
The black dotted curve sets a baseline
and is the bare band case. The red dashed, which is for comparison,
is $[1-d{\rm Re}\Sigma(\omega)/d\omega] {\rm sgn}(\omega+\omega_d)$. 
(b) Same as for (a) but with $\mu_0=500$ meV. (c) The absolute height of
the phonon peaks at about $\omega=\omega_E$ (solid blue dots) and 
$-(\mu_0+\omega_E)$ (open red circles) along with the $\lambda$ variation
with $\mu_0$, the latter indicated as the red solid curve using the righthand axis .
}
\label{fig3}
\end{figure}

\begin{figure}[ht]
\begin{picture}(250,200)
\leavevmode\centering\includegraphics{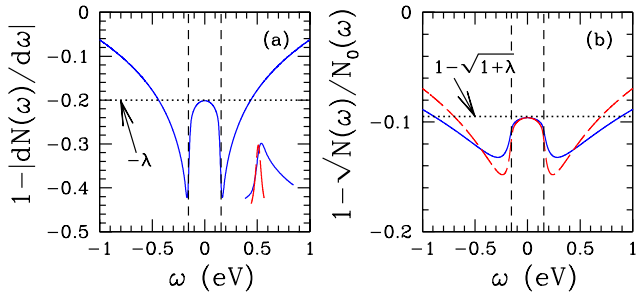}
\end{picture}
\vskip -100pt
\caption{(Color online) (a) $1-|dN(\omega)/d\omega|$ vs $\omega$ for $\mu_0=0$
and $\omega_E=155$ meV (solid
  blue curve). The
  inset on the lower right
compares the phonon region with the input electron-phonon spectral
density
(red-dashed curve). (b) $1-\sqrt{N(\omega)/N_0(\omega)}$ vs $\omega$
(solid blue curve) compared with the result using the procedure of Li
et al. (long-dashed red curve).
}
\label{fig4}
\end{figure}

Phonon structure in $N(\omega)$ can be brought out through
differentiation.
Results for $dN(\omega)/d\omega$ vs $\omega$ are given in
Fig.~\ref{fig3}
as the solid blue curves, where $N(\omega)$ is 
normalized by $N_0$. Frame (a) is for $\mu_0=150$ meV
and (b) is for $500$ meV. The vertical drop where 
$dN(\omega)/d\omega$ goes from positive to negative
is at the Dirac point of the interacting system. Comparison with the
bare band case, the dotted black  line, shows a small shift
of 
the position of the Dirac point between bare and dressed case.
The bare case provides a useful reference line about which the
effects
of the electron-phonon interaction are easily seen. Besides the phonon
structures
at $\omega=\pm\omega_E,-(\omega_E+\mu_0)$, we note that the height of
the
curve above one at $\omega=0$ gives the value of $\lambda$ directly which
increases significantly with increasing value of chemical potential
[as shown by the red solid curve in (c)]. 
The red dashed
line
is included for comparison and gives $-d{\rm
  Re}\Sigma(\omega)/d\omega$. There are some differences between these
two
sets of results but we can conclude that all qualitative features seen
in the DOS curves can be seen in the ${\rm Re}\Sigma(\omega)$.
This is not to say that the imaginary part of $\Sigma(\omega)$ plays
no significant role. In frame (b), we see clearly that the jump
at the Dirac point energy is no longer vertical but exhibits some
smearing.
This can be traced to the behavior about the Dirac point in the DOS
shown in Fig.~\ref{fig2}(b). The DOS
no longer goes to zero at this point ($\omega_d$) but rather has a
minimum
about which it rises as a quadratic $(\omega-\omega_d)^2$, 
seen in experiment\cite{wang}. We can show
that for  $|(\omega-\omega_d)Z|\ll\Gamma$,
\begin{equation}
\frac{N(\omega)}{N_0}=\frac{2\Gamma}{\pi}\ln\biggl|\frac{W_C}{\Gamma}\biggr|+
\frac{(\omega-\omega_d)^2Z^2}{\pi\Gamma},
\label{eq:dosdirac}
\end{equation}
with $Z\equiv 1-[d{\rm Re}\Sigma(\omega)/d\omega]_{\omega=\omega_d}$
and  $\Gamma\equiv -[{\rm Im}\Sigma(\omega)]_{\omega=\omega_d}$,
which shows the lifting of the Dirac point and its conversion from
linear to quadratic in $(\omega-\omega_d)$. This immediately leads to
the
smearing at Dirac point noted in the blue curve of Fig.~\ref{fig3}(b).
In Fig.~\ref{fig3}(c), we plot the absolute value
of
the  height
of the phonon structures as a function  of $\mu_0$ for the $\alpha^2F(\nu)$
spectrum used here, a truncated Lorentzian (see in the inset
of Fig.~\ref{fig4}, long-dashed red curve). While the height of the
phonon peak at $\omega=-(\omega_E+\mu_0)$ hardly changes with doping $(\mu_0)$
the other two peaks do, note the curve for $\omega=\omega_E$.
We have not plotted the peak height for $\omega=-\omega_E$ as for $\mu_0<\omega_E$
it is similar to the result for $\omega=\omega_E$ and for $\mu_0>\omega_E$
it becomes ambiguous.
These predictions provide verifiable tests that observed structures are 
indeed due to phonons. They also show how the increase in the DOS
at the Fermi surface with increasing doping is reflected in larger coupling to
the phonons.

In Fig.~\ref{fig4}(a), we show results for $(1-|dN(\omega)/d\omega|)$ in
the specific case of $\mu_0=0$ which is close to the case recently observed in
the scanning tunneling miscroscopy (STM) results of Li et al.\cite{li} To
conform
with those experiments, we have used $\omega_E=155$ meV, with 
$\omega=\pm\omega_E$ 
shown as the vertical black dashed lines. The
horizontal
black dotted line through the local maximum of the solid blue curve at
$\omega=0$ identifies the value of $\lambda$ which we took to be 0.2.
A second important feature is the phonon structure which reflects the
underlying $\alpha^2F(\nu)$ spectrum used. It is shown   in the inset
as the long dashed red curve where it is scaled down and
 compared with the absolute
value of the blue
curve about $\omega=155$ meV. While there is some agreement, the two curves have different
profiles with the blue solid one much broader than the red long-dashed one.
It is clear that such a plot is very useful in identifying phonon
structure, i.e., not just the value of the mass enhancement factor $\lambda$
involved but also the position of the peaks in $\alpha^2F(\nu)$ and
their
strength. In experiments, it may be more desirable not to
differentiate.
In Fig.~\ref{fig4}(b), we show as the solid blue curve
a different quantity $1-\sqrt{N(\omega)/N_0(\omega)}$, where
$N_0(\omega)$
is the bare band density of states. In this quantity, the value of the
local
maximum at $\omega=0$ gives $1-\sqrt{1+\lambda}=-0.095$ rather than the
$-\lambda$
of frame (a). We also note that the phonon structures at
$\omega=\pm\omega_E$
are not as sharp, however, some signature of a sharp peak in the
$\alpha^2F(\nu)$
used remains. The long-dashed red curve is for comparison and
represents
the quantity that was used by Li et al.\cite{li} in their analysis of their STM
data. They use a definition of an effective Fermi velocity dependent
on $E$ based on an integration of their conductance. 
They define $v_F^{\rm eff}=dE/\hbar dk$ with 
$k=\pm|\int_{\omega_d}^EN(\omega)d\omega|^{1/2}$, where $\omega_d$ is
the energy locating the Dirac point. Like the blue solid curve, the
maximum
at $\omega=0$ provides $1-\sqrt{1+\lambda}$ and the phonon structures
at $\omega=\pm\omega_E$ are clearly seen. 
The Li et al.\cite{li} estimate of $\lambda=0.26$ is close to the 0.3 value
from ARPES.\cite{bostwick} Our own estimate based on the STM data is somewhat higher, but carries considerable uncertainty because of the experimental noise.
An important point to note between Fig.~\ref{fig4}(a) and (b) is to reiterate
that the many body renormalizations correct $N(\omega)$ by a $(1+\lambda)$
factor. Assuming $|\epsilon|/v_F^2\to|\epsilon|/v_F^{*2}$ would
over estimate the correction by an additional factor of $(1+\lambda)$. 

In contrast to the standard 
expectation in wide band metals with nearly constant
DOS on the phonon energy scale, phonon structure does appear prominently
in the DOS of graphene and this can be used as a new spectroscopy for
determining electron-phonon coupling. We find that the mass enhancement
parameter $\lambda$ can be extracted directly from the data around the
Fermi energy. In addition, three prominent peaks are identified at
$\omega=\pm\omega_E$ and $\omega=-(\omega_E+\mu_0)$ associated
with each Einstein mode. The size of these additional structures increases with
increasing doping as does the mass enhancement $\lambda$. These increases
reflect the increase in the underlying DOS. In this sense,
graphene provides a new laboratory in which to study variations in 
electron-phonon coupling with changing carrier concentration. 

We thank Sergei Sharapov for valuable assistance and insight,
and Eva Andrei and Guohong Li for discussion about their work.
This research has been supported by NSERC of Canada (E.J.N. and J.P.C.)
and by the Canadian Institute for Advanced Research (CIFAR) (J.P.C.).

\end{document}